\documentclass[review]{elsarticle}
\usepackage{lineno,hyperref,amssymb,amsmath,booktabs,subfigure}
\usepackage{hyperref}
\usepackage[dvipsnames]{xcolor}
\biboptions{numbers,sort&compress}

\journal{Electrochimica Acta}

\newcommand{\bi}{ \boldsymbol}

\begin{document}
\begin{frontmatter}

\title{Modeling discontinuous potential distributions using
the finite volume method, and application to liquid
metal batteries}

\author[hzdr,mit]{Norbert Weber}
\author[hzdr]{Steffen Landgraf}
\author[mit,porto,pakistan]{Kashif Mushtaq}
\author[hzdr]{Michael Nimtz}
\author[hzdr]{Paolo Personnettaz}
\author[hzdr]{Tom Weier}
\author[mit]{Ji Zhao}
\author[mit]{Donald Sadoway}

\address[hzdr]{Helmholtz-Zentrum Dresden -- Rossendorf, Bautzner
  Landstr.\ 400, 01328 Dresden, Germany}
\address[mit]{Department of Materials Science and Engineering, Massachusetts Institute of Technology, 77 Massachusetts
  Avenue, Cambridge, MA 02139-4307, United States}
\address[porto]{LEPABE, Department of Chemical Engineering, Faculty of Engineering, University of Porto, Portugal}
\address[pakistan]{Department of Mechanical Engineering, School of Mechanical and Manufacturing Engineering, National University of Sciences and Technology, Islamabad, Pakistan}

\begin{abstract}
The electrical potential in a battery jumps at each
electrode-electrolyte interface. We present a model for computing
three-dimensional current and potential distributions, which accounts 
for such internal voltage jumps. Within the framework of the finite
volume method we discretize the Laplace and gradient operators such
that they account for internal jump boundary conditions. After
implementing a simple battery model in OpenFOAM we validate it using
an analytical test case, and show its capabilities by simulating the
current distribution and discharge curve of a Li$||$Bi liquid metal
battery.
\end{abstract}


\end{frontmatter}


\section{Introduction}
The objective of this article is twofold: we want to model both, the
current distribution in, and charge-discharge curves of electrochemical 
cells. For good overviews on the subject, see
\cite{Prentice1982,Dukovic1990,Weber2004}. As a first step, we will
discuss only modeling of the time-dependent cell potential.

\subsection{Battery voltage, overpotentials and polarization curve}
Most battery models do not describe the distribution of current
and potential. Instead, they provide only the total cell voltage,
which can be simply calculated as the difference of the open circuit
potential and the sum of all overvoltages
\cite{Vetter1967,Pierini1976,Newman2004,Ismail2014,Beale2016}.
We are here especially interested in the ohmic overpotential. In many
batteries, the electrolyte has by far the highest resistance.  In such
cases the ohmic losses can be calculated 
analytically or by potential theory,
i.e., by solving a Laplace equation in the electrolyte 
\cite{Deconinck1985,Camprubi2011,Colli2017}. More advanced models are
used for porous electrodes, but rely on the same Laplace equation
\cite{Newman1962,Newman1975,Doyle1995,Fuller1994,Paxton1997}. Even if
the full potential distribution is computed in the electrolyte, all these
models use in the end only a \emph{scalar} voltage loss to compute the
cell potential. They are therefore limited in many ways -- and can not
account for effects such as
the potential drop in poorly conducting electrodes (e.g. Se, S
\cite{Hueso2013}), complicated geometries where much current flows
only through a part of the electrolyte, local activation
overpotentials, or liquid electrodes where diverging current drives
convection \cite{Ashour2017a,Weber2018}.

\begin{figure}[h!]
\centering
\includegraphics[width=0.7\textwidth]{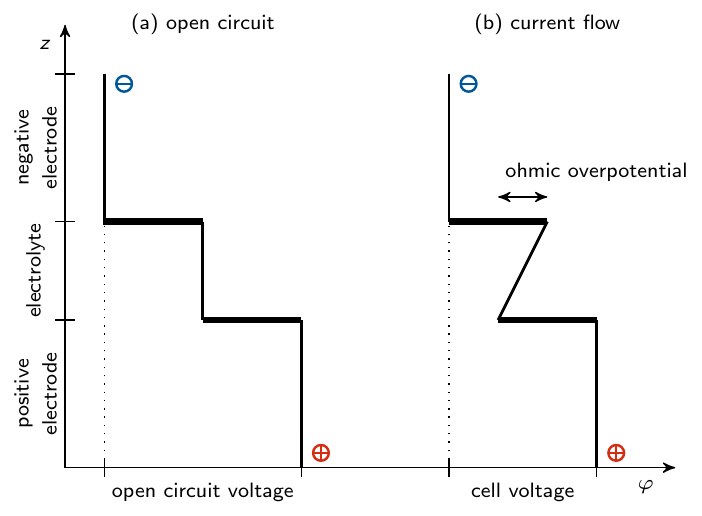}
\caption{Schematic voltage profile in an electrochemical cell with open (a)
  and closed (b) circuit at discharge. For an excellent discussion of such profiles, see
\cite{Vetter1967,Wang1998,Bard2001,Luck2016}.} \label{f:voltageProfile}
\end{figure}

In order to account for all these effects, we need to know the
potential distribution in the battery. Fig.\,\ref{f:voltageProfile} shows a
one-dimensional example -- an example
which is strongly simplified. It relies on the macroscopic approach,
i.e., we do not resolve the electrochemical double layer, and consider
all phases to be electrically neutral\footnote{For
  simulating alternating   current response of electrochemical cells
  it will be necessary to   consider the effects of the
  electrochemical double layer. For more   information on the subject,
  see, e.g., \cite{Biesheuvel2009}.} \cite{Newman1975,Schalkwijk2002,Weber2004}. Within these assumptions,
the potential jumps at both electrode-electrolyte interfaces according
to the Nernst equation. As illustrated in
fig.\,\ref{f:voltageProfile}a the cell voltage is simply the
difference of both potential jumps. During operation of the
electrochemical cell, the potential jumps are directly reduced by
concentration and activation overpotentials. Additionally, the cell
potential decreases by the ohmic overpotential, as illustrated in
fig.\,\ref{f:voltageProfile}b.

\subsection{Modeling current and potential distributions}
After describing one-dimensional potential profiles in the previous
section, we will proceed with a discussion of 3-dimensional
modeling. Within continuum mechanics, potential distributions can be
obtained by solving a Laplace equation
\begin{equation}\label{eqn:laplace}
  \nabla\cdot\sigma\nabla\varphi = 0,
\end{equation}
where $\varphi$ denotes the electric potential and $\sigma$ the ionic
or electronic conductivity. This simple approach is only fully correct
if there are no concentration gradients in the electrolyte
\cite{Klingert1964,Newman2004} and if double layer charging is
neglected \cite{Weber2004}. The challenge of solving equation
\ref{eqn:laplace} is the potential jump between electrolyte and
electrodes. Generally, two solutions exist: the single- and the
multi-potential approaches \cite{Beale2007}. The latter relies on the idea of defining 
separate ionic and electronic potentials \cite{Smith2007,Feinauer2018},
i.e., several Laplace equations are solved. The offset,
i.e., the jump between the potentials, is typically defined by a
volumetric source term in an envelope layer near the interface \cite{Kaoui2018}. Such a
model is ideally suited for porous electrodes, such as the catalyst
layer in a fuel cell. The porous electrode and electrolyte are treated
as superimposed continua \cite{Newman1962,Smith2007} of finite
thickness \cite{Newman2004}; a charge leaving the solid matrix must
enter the pore liquid \cite{Gu1997}. The superficial charge transfer --
described by Faraday's law -- needs to be transferred into only a
volumetric source term \cite{Newman1962,Schalkwijk2002}. Such
multi-potential porous electrode models are extremely popular -- for
more details, see especially
\cite{Euler1960,Newman1962,Newman1975,Milshtein2017,Gu1997}.
Please note that many
researchers exclude the potential jump due to the Nernst potential,
but model only the jump due to the activation overpotential.

Compared to the above described porous model, the single-potential
approach is less common. It uses only one single potential field
for the whole  cell. The potential jump
is not implemented as a source term, but as an internal jump boundary
condition. Such models are also known as ``interface models'' because
they assume the interface to be infinitely thin \cite{Weber2004}. We
will implement such a model and describe in the following how the
equations are set up and how they are solved.

\subsection{Solving coupled equation systems with internal jumps}
As described in the previous section, the electric potential between
electrodes, electrolyte and conductors needs to be coupled. We will
denote this by the term ``region coupling''. The voltage of
 electrochemical cells depends, among others, on temperature and
concentration. Each of these fields is described by its own equation --
and all of them are coupled. We will denote that as 
``equation coupling''. Newman's original method for coupling different
equations \cite{Newman1968} is
based on block-matrices, i.e., all equations are discretized in one matrix
and solved in one step \cite{Newman1967,White1978,Varga2000}. This
method was slightly extended \cite{Kimble1990,Nguyen1987,Mao1994,Preisig1990}
and used for many different applications. For a good overview about
the method, see \cite{Botte2000}. An alternative to block-coupling is
a segregated solver: it solves each equation separately.
As the equations are coupled, this process must be
repeated iteratively until convergence is reached. On the one hand, the
block-coupled single-matrix approach is surely better suited for
highly coupled equations, and maybe for strongly non-linear equations
\cite{Beale2018}. On the other hand, the segregated approach is easier
to implement and needs much less computer memory. Both approaches have
advantages and drawbacks -- there is not a single, and perfect solution
for equation coupling.

Similarly, region coupling can be obtained by solving the potential on
one global mesh, or by solving a separate equation for each conductor. The
latter is well known in OpenFOAM. Using appropriate boundary
conditions, a potential jump at the interface can easily be implemented
\cite{Takayama2011,Colli2018}. However, a relatively time consuming 
iteration between all regions is necessary. This can be extremely 
slow \cite{Weber2017b}, because the regions are coupled only at the 
interface (and not in the volume). Solving the
potential in all regions in a coupled way in one matrix is definitely
the better way. For possible implementations of internal potential
boundary conditions, see \cite{White1984,Dimpault-Darcy1988,Fan1991}.

Finally, the equation system can be set-up and solved using the
finite-difference (FDM), finite-element (FEM), boundary
element or finite-volume method (FVM)
\cite{Deconinck1985,Dukovic1990}. During the first years, the FDM was
surely the most widely used approach for modeling potential problems
\cite{Newman2004}. Today, commercial codes such as COMSOL
\cite{Ziegler2004} (FEM), STAR-CCM+ \cite{Ju2004}
(FVM) and Ansys Fluent \cite{Karimi-Sibaki2018a}
(FVM) also allow for solving potential problems with internal
jumps. However, their solution algorithm is not always published.

\subsection{Our approach}
We will model potential distributions with internal jumps using a
single-potential approach. The set of equations will be solved in a
segregated manner, while a block-matrix will be used for region
coupling of the electric potential. We further use a multi-mesh
approach \cite{Beale2016,Weber2017b}. This means, we solve certain
variables (as the potential) on a global mesh, but others (such as
concentration) only in the electrodes. The equations are discretized
using the finite volume method, and the model is implemented in the
open source CFD library OpenFOAM v1806 \cite{Weller1998}.

\section{Model}
We present in this section the most simple model for the potential
distribution in a concentration cell. With little effort, it can 
be extended to arbitrary electrochemical cells. We use the
parent child-mesh technique, i.e., we provide one global mesh for the
full geometry and a second mesh for the positive electrode only. Where
necessary, we map variables between both meshes. The following
simplifications apply:
\begin{itemize}
\item fluid flow, heat transfer and variation of the layer thickness are neglected
\item the charge transfer overpotential is neglected
\item the electrochemical double layer is not resolved; we assume
    discrete potential jumps at the interfaces
\item concentration gradients in the electrolyte are neglected
  \cite{Vallet1983} 
\end{itemize}
In a first step we solve the concentration $c$ in the positive electrode
as
\begin{equation}
\frac{\partial}{\partial t} c = \nabla\cdot(D\nabla c),
\end{equation}
where $D$ denotes the diffusion coefficient. We apply zero flux
boundary conditions at all interfaces except the electrode-electrolyte
boundary. Here we set the concentration gradient corresponding to the
current density $\bi J$ as
\begin{equation}
\nabla_n c = - \frac{\bi J}{z F D}\bi n,
\end{equation}
where $\nabla_n$ denotes the interface-normal gradient, $z$ the number of
electrons, $F$ the Faraday constant and $\bi n$ the interface normal vector.

The local concentration at the electrode-electrolyte interface
determines the jump in potential. The latter could be computed
from the Nernst equation using measured activities. However, for
simplicity and to avoid numerical instabilities \cite{Beale2018}, we
directly fit the measured open circuit voltage.
 We save this potential jump at the corresponding
face centers and map it to the parent mesh. Please note that
the potential jump is applied only at the interface between electrolyte
and the positive electrode (and not at the negative electrode).

On the global mesh we solve a Laplace equation for the electric
potential $\varphi$ as
\begin{equation}\label{eqn:Laplace}
\nabla \cdot\sigma\nabla\varphi = 0,
\end{equation}
where $\sigma$ denotes the electrical conductivity. Finally, the current
density is computed as 
\begin{equation}\label{eqn:grad}
  \bi J = \sigma\nabla\varphi,
\end{equation}
and mapped to the electrode-electrolyte
interface where it is needed to compute the boundary condition for the
concentration. The potential jump at the electrolyte-positive electrode interface
is accounted for when discretizing the gradient and Laplace operator
as described in the next section.

We discretize the equations using the implicit Euler scheme for time
and second order schemes for all spatial terms. The potential is
solved using a PCG, and the concentration by a multigrid solver
\cite{Ferziger1996}.

\section{Discretization of jump condition}\label{s:discretisation}
Solving the electrical potential on one single mesh, one needs to
account for the jumping potential in two terms. The first is the Laplace
equation \ref{eqn:Laplace} where the potential jump will appear in
form of an additional source term. Secondly, when computing the
current density by equation \ref{eqn:grad}, both the potential jump
and the discrete change of conductivity has to be observed.

\subsection{Laplace operator}
Within the finite volume method, the Laplace operator can be
discretized using the Gauss theorem as \cite{Jasak1996}
\begin{equation}
\nabla\cdot\sigma\nabla\varphi = \sum_f \sigma_f\bi S(\nabla\varphi)_f, 
\end{equation}
with $\bi S$ denoting the face area vector and $(\nabla\varphi)_f$ the
gradient at the face. The face conductivity $\sigma_f$ is consistently
discretized from the cell centered values\footnote{We use the word ``cell''
in section \ref{s:discretisation} in the sense of grid cell or
control volume.} using harmonic interpolation
\cite{Weber2017b}.
Denoting the potential in the owner and neighbor cell of face $f$ by
$\varphi_\mathrm{P}$ and $\varphi_\mathrm{N}$ we can also write 
\begin{equation}
\nabla\cdot\sigma\nabla\varphi = \sum_f \bi S \sigma_f(\nabla\varphi)_f = \sum_f |\bi S| \sigma_f\frac{\varphi_\mathrm{N}-\varphi_\mathrm{P} + \Delta\varphi}{|\bi d|},
\end{equation}
where $\Delta\varphi$ denotes the potential jump at the
interface and $\bi d$ the vector connecting both cell centers. Writing the matrix equation as
\begin{equation}
  A_d + A_{o} = S
\end{equation}
we find the off-diagonal coefficients as
\begin{equation}
  A_{o} = \frac{\sigma_f|\bi S|}{|\bi d|},
\end{equation}
and the diagonal coefficients as
\begin{equation}
A_d = -\sum_n \frac{\sigma_f|\bi S|}{|\bi d|} = -\sum_r A_{r,o},
\end{equation}
where the index $r$ denotes related cells, i.e., cells which share a
common face\footnote{For details, see \url{https://openfoamwiki.net/index.php/OpenFOAM_guide/Matrices_in_OpenFOAM}.}. Finally, the source term is
\begin{equation}
S = \sum_f -\frac{\sigma_f|\bi S| \Delta\varphi}{|\bi d|}.
\end{equation}
The potential jump $\Delta\varphi$ will appear only at the electrode-electrolyte interface;
for all grid cells not touching the interface the source term will
therefore be zero.

The above described procedure applies perfectly to orthogonal
meshes. It can be easily extended to arbitrary polyhedral control
volumes using the overrelaxed correction approach
\cite{Hill2018,Westhuizen2013,Jasak1996,Ferziger1996,Demirdzic2015}.

\subsection{Gradient operator}
Using the Gauss theorem, the gradient of the electric potential is
discretized as \cite{Jasak1996}
\begin{equation}
\nabla\varphi = \frac{1}{V}\sum_f\bi S\varphi_f,
\end{equation}
where $V$ denotes the cell volume and $\varphi_f$ the potential at a
face. The latter is determined from the potential of the owner
(P) and neighbor (N) cell using two continuity conditions. Firstly,
we assume the potential itself to be continuous (or jumping) over the
face as
\begin{equation}
  \varphi_\mathrm{wN} = \varphi_\mathrm{wP} + \Delta\varphi,
\end{equation}
and secondly we ensure continuity of normal currents by
\begin{equation}
\sigma_\mathrm{N}\nabla\varphi_\mathrm{N}\cdot\bi n_\mathrm{N} = \sigma_\mathrm{P}\nabla\varphi_\mathrm{P}\cdot\bi n_\mathrm{P},
\end{equation}
where $\varphi_\mathrm{wP}$ and $\varphi_\mathrm{wN}$ denote the face
potential in the owner and neighbor cell and $\Delta\varphi$ the jump
between both; $\sigma_\mathrm{P}$ and $\sigma_\mathrm{N}$ denote (cell centered)
conductivities, $\bi n_\mathrm{P}$ and $\bi n_\mathrm{N}$ the face normal vectors and
$\varphi_\mathrm{N}$ and $\varphi_\mathrm{P}$ the cell values of the potential. Combining
both above conditions leads to the face potential in the owner cell as
\begin{equation}
\varphi_\mathrm{wP} = f\cdot(\varphi_\mathrm{N}-\Delta\varphi) + (1-f)\cdot\varphi_\mathrm{P}
\end{equation}
with
\begin{equation}
f = \frac{\delta_\mathrm{P}\cdot\sigma_\mathrm{N}}{\delta_\mathrm{N}\sigma_\mathrm{P}+\delta_\mathrm{P}\sigma_\mathrm{N}}.
\end{equation}
Here, $\delta_\mathrm{P}$ and $\delta_\mathrm{N}$ denote the distance
between face and cell center for the owner and neighbor cell,
respectively.

\section{Validation}
\begin{figure}[h!]
\centering
\includegraphics[width=0.5\textwidth]{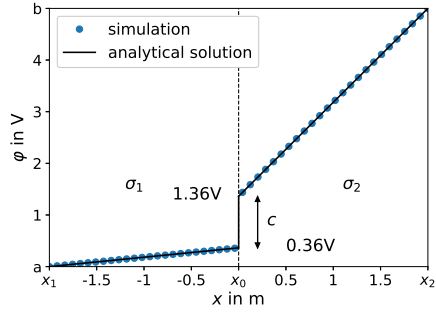}
\caption{Electric potential distribution along a line for the
  one-dimensional test case. The potential at the boundaries (a, b) as
  well as the voltage jump (c) and the two conductivities ($\sigma_1$, $\sigma_2$) are provided as initial
  conditions.}\label{f:analyticalPhi}
\end{figure}

We use a simple analytical test case to validate our model. A 
1-dimensional bar is made of two materials of different conductivity,
and an artificial potential jump of 1\,V is applied between both. Further, a constant
potential is applied at both ends of the bar, such that a potential profile
as illustrated in fig.\,\ref{f:analyticalPhi} develops. The 
following parameters are used:
\begin{align}
  a &= 0\,\mathrm{V}, \quad
  b = 5\,\mathrm{V}, \quad
  c = 1\,\mathrm{V},\nonumber\\
  x_0 &=0\,\mathrm{m}, \quad
  x_1 =-2\,\mathrm{m}, \quad
  x_2 =2\,\mathrm{m},\nonumber\\
  \sigma_1 &=10\,\mathrm{S/m}, \quad
  \sigma_2 =1\,\mathrm{S/m}.\nonumber
\end{align}
The analytical solution predicts the potential profile as two lines:
\begin{align}
  \varphi_1(x) = \alpha x + \beta,\\
  \varphi_2(x) = \gamma x + \varepsilon.
\end{align}
with
\begin{align}
\alpha &= \sigma_2\frac{b-c-a}{\sigma_2(h_0-h_1)-\sigma_1(h_0-h_2)},\\
\beta &= a - \alpha h_1,\quad
\gamma = \alpha\frac{\sigma_1}{\sigma_2},\quad
\varepsilon = b- \gamma h_2.
\end{align}

The simulated electric potential in fig.\,\ref{f:analyticalPhi} shows
the jump as expected, and fits perfectly to the analytical
solution. Moreover, the current density (not shown here) is continuous
over the interface, and fits again to the analytical solution.

\section{Application to a Li$||$Bi liquid metal battery}
\begin{figure}[h!]
\centering
\includegraphics[width=0.5\textwidth]{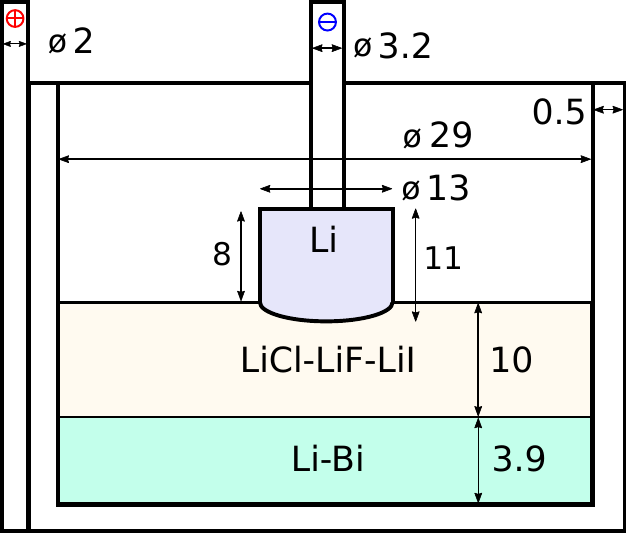}
\caption{Experimental Li$||$Bi cell. The vessel is made of tantalum,
  the wires of copper. The lithium metal is contained in a spiral made
  of nickel. The Li-Bi layer thickness corresponds to a Li molar fraction of 0.236.}\label{f:wunderzelle}
\end{figure}

Liquid metal batteries are discussed as cheap stationary
energy storage for fluctuating renewable energies 
\cite{Kim2013b,Wang2014}. We simulate a Li$||$Bi liquid metal battery \cite{Foster1964,Ning2015} in 3D (for a 1D model, see \cite{Newhouse2014}) -- 
and compare with measured data. The concentration cell consists of
a Bi positive electrode (0.1\,mol), a LiF-LiCl-LiI
molten salt electrolyte and a Li
negative electrode \cite{Weier2017}. The lithium is contained in a nickel spiral, and
the vessel is made of tantalum to avoid as far as possible any reaction
with the active materials and the electrolyte.
Fig.\,\ref{f:wunderzelle} shows the setup, and
tab.\,\ref{t:material} gives the material properties. The cell is
heated from below such that the temperature at the bismuth-electrolyte
interface stays at approximately 460$^\circ$C. Ultra-dry LiI is vacuum dried for 12\,h
 while being heated step wise from 100 to
300$^\circ$C. LiF-LiCl is vacuum dried for 12\,h at
500$^\circ$. Finally, LiF-LiCl-LiI is mixed in eutectic composition
\cite{Masset2006b} and filtered through a
quartz frit. The cell is cycled at 1\,A with a cycle length of
10\,min.

\begin{figure}[h!]
\centering
\subfigure[]{\includegraphics[width=0.4\textwidth]{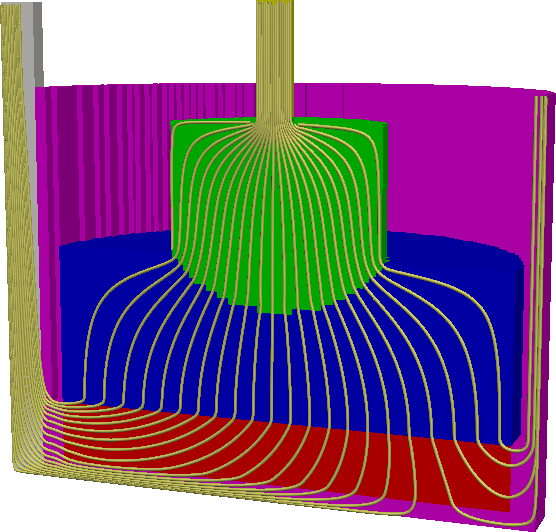}}
\subfigure[]{\includegraphics[width=0.45\textwidth]{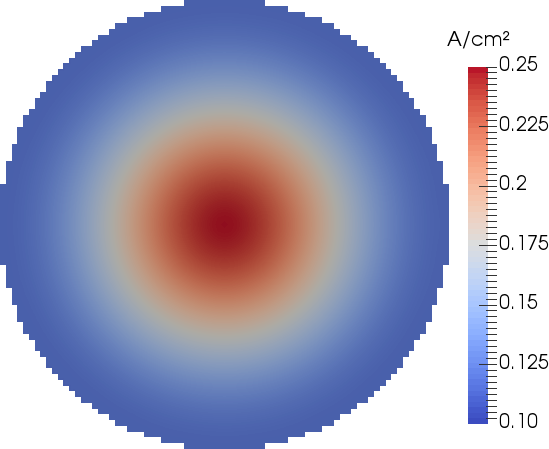}}

\subfigure[]{\includegraphics[width=0.45\textwidth]{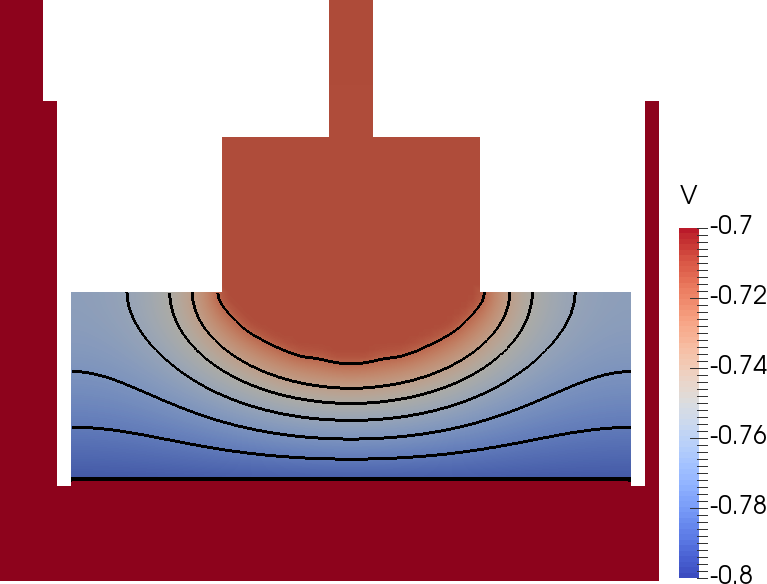}}
\subfigure[]{\includegraphics[width=0.45\textwidth]{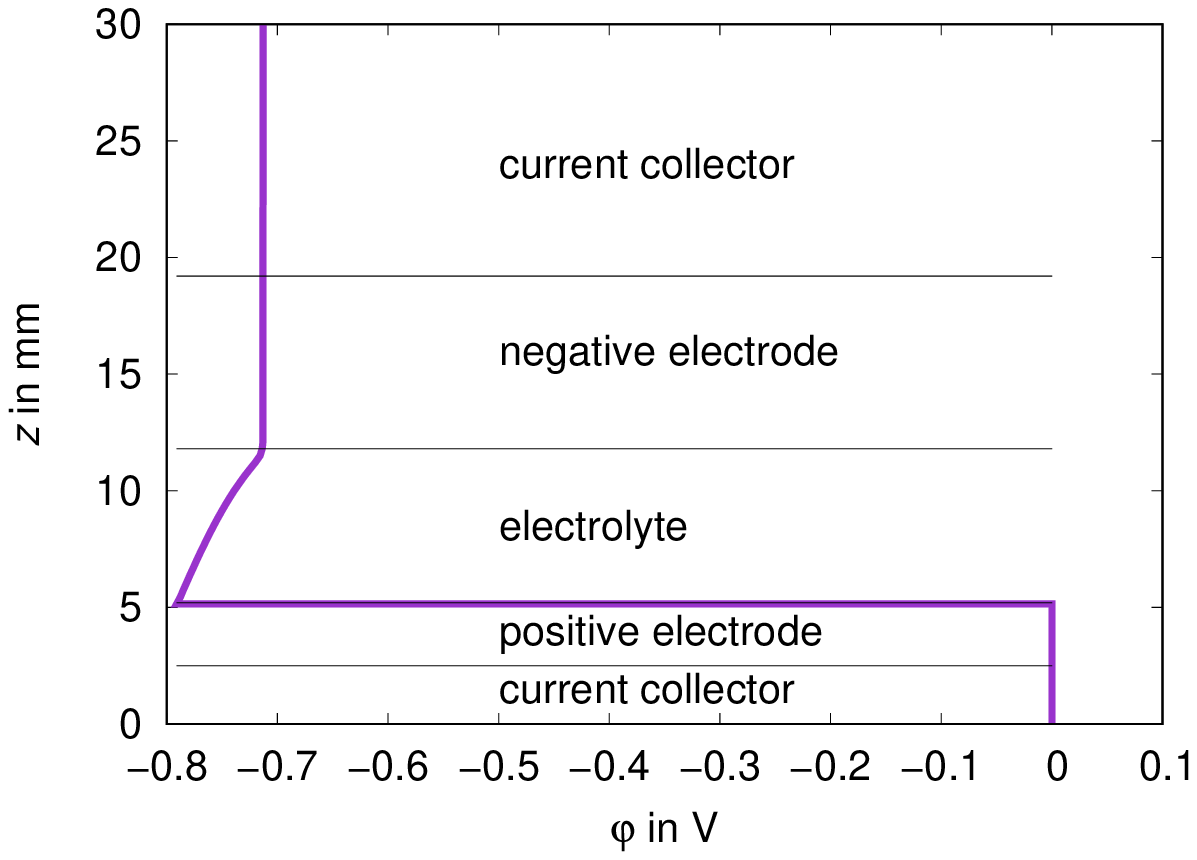}}

\subfigure[]{\includegraphics[width=0.4\textwidth]{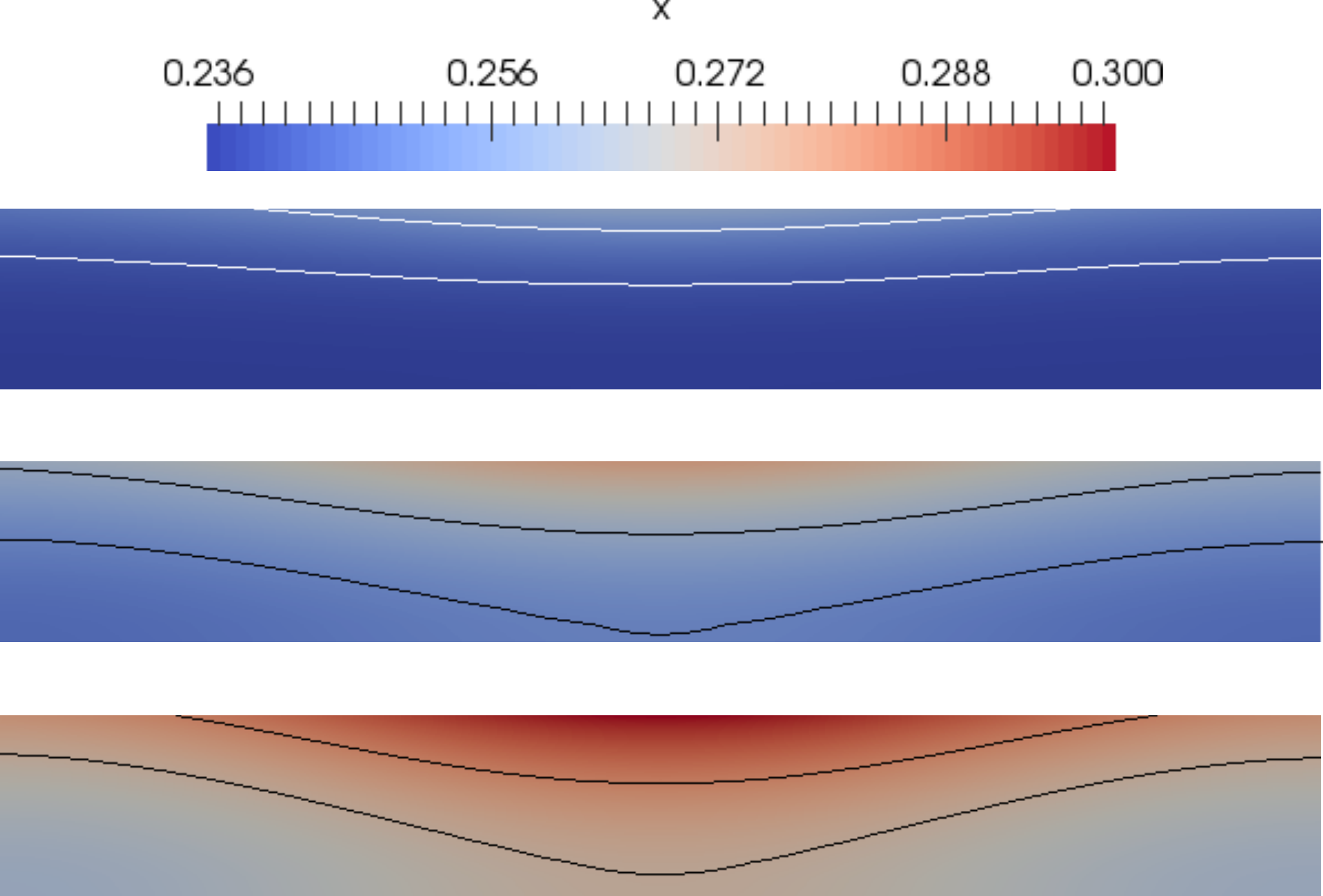}}
\subfigure[]{\includegraphics[width=0.45\textwidth]{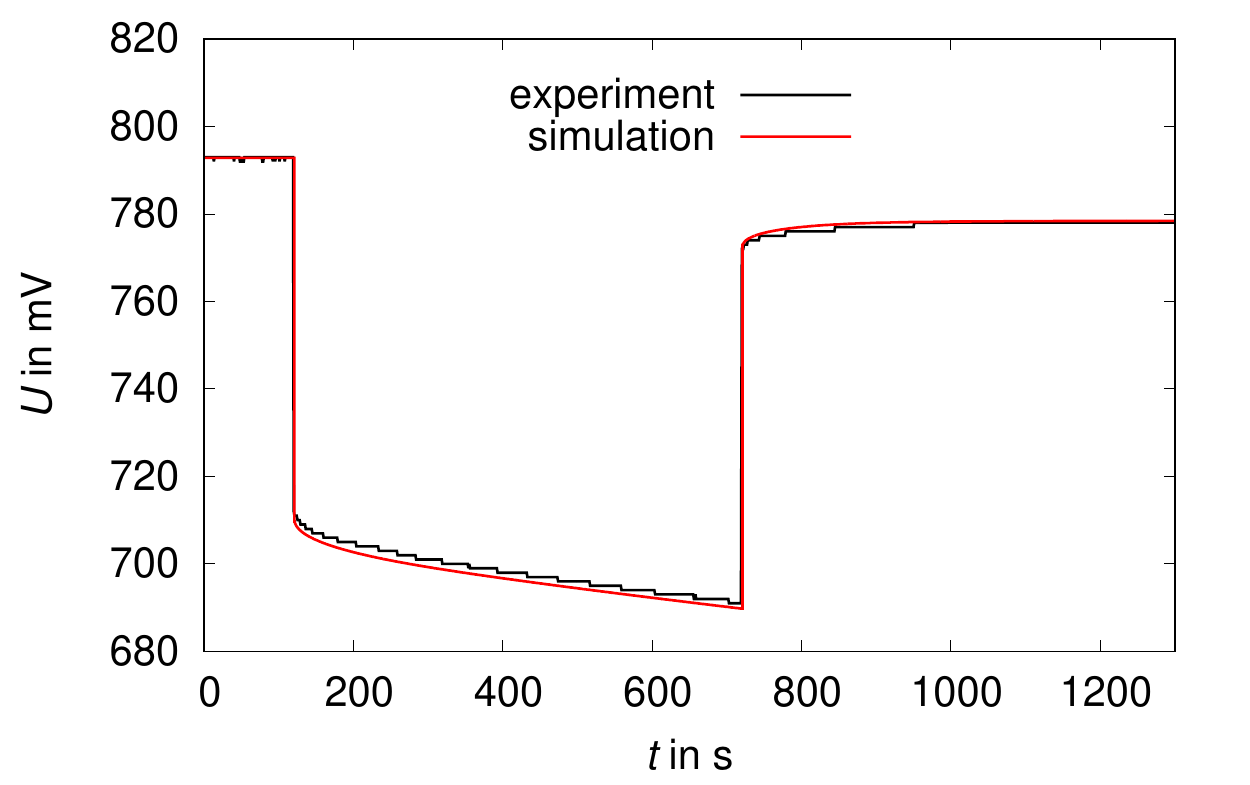}}
\caption{Grid with current distribution (a), current distribution at the
electrolyte-positive electrode interface (b) and electric potential (c)
in a Li$||$Bi liquid metal battery with a molar fraction of Li in Bi of
$x=0.236$, as well as the vertical potential distribution along the axis of the cell
(d). The discharge curve (f) matches fairly well between experiment
and simulation; (e) shows the corresponding Li-molar fraction in the
positive electrode after 1, 5 and 10 minutes.} 
\label{f:experiment}
\end{figure}

The numerical model is simplified in three ways. Firstly, we insert 
a very thin gap artificially between electrolyte and vessel, because
no current is allowed to flow there. Secondly, we model the negative
electrode as it would consist of pure lithium. This is justified 
because the electric conductivities of nickel and lithium are very 
similar. Finally, we do not know the exact shape of the electrolyte-negative
electrode interface -- the exact immersion depth shown in fig.\,\ref{f:wunderzelle} is
therefore an assumption. The numerical representation of the cell is shown in fig.\,\ref{f:experiment}a: 100 grid cells
are used on the diameter of the FVM mesh, and the electrolyte-positive electrode
interface is strongly refined. The potential is set to 0\,V at the
outer cable while a Neumann boundary condition corresponding to 1\,A
is applied at the negative contact. The open circuit potential
at 460$^\circ$C is fitted using measurement values \cite{Gasior1994}
as
\begin{equation}
E / \mathrm{V} = \frac{0.19}{x+0.41} + 0.5
\end{equation}
for the interval of molar fraction $0.1< x< 0.3$. The density of the Li-Bi alloy is computed using Vegard's law \cite{Vegard1921,Fazio2015}.
 
\begin{table}[h!]
\caption{Material properties at 460$^\circ$C
  \cite{Zinkle1998,Sobolev2010,Masset2007,Gale2004,Personnettaz2018a,crc2005}.}
\centering\label{t:material}
\begin{tabular}{lrrrrrr}
\toprule
\multicolumn{1}{c}{property} & \multicolumn{1}{c}{Li} &
                                                        \multicolumn{1}{c}{Bi}
  & \multicolumn{1}{c}{salt} & \multicolumn{1}{c}{Cu} &
                                                        \multicolumn{1}{c}{Ta} & \multicolumn{1}{c}{Ni}\\
\midrule
$\rho$ in kg/m$^3$  & 490 &  9831 &2690& 2800 & \\
$\sigma$ in S/m & 2.7$\cdot 10^6$ & 7.2$\cdot 10^5$ &271 & $58\cdot10^6$ & $2.9\cdot10^6$ & $3\cdot10^6$\\
\bottomrule
\end{tabular}\end{table}

Fig.\,\ref{f:experiment}c illustrates the electric potential field, and
fig.\,\ref{f:experiment}d the profile along the axis. We clearly see
the potential jump corresponding to the open circuit voltage.
 The current density, as illustrated in fig.\,\ref{f:experiment}a shows
 only an asymmetry at the bottom. Despite all current needs to flow to the
 lateral cable, the current distribution in the electrolyte is almost
 symmetric. Finally, fig.\,\ref{f:experiment}f shows the 10\,min
 discharge cycle of 1\,A, starting with a molar fraction of Li in Bi of
 $x=0.236$. Experiment and simulation fit very well. The corresponding
 Li molar fraction in Bi after 1, 5 and 10 minutes is shown in
 fig.\,\ref{f:experiment}e. As expected from the current distribution in fig.\,\ref{f:experiment}b,
 we find more Li in the center of the positive electrode. 
In reality, this could
 finally lead to the formation of intermetallic phases there.

\section{Summary and outlook}
We have discussed different approaches for modeling the cell voltage, as
well as the  macroscopic potential and current distribution in
electrochemical cells. Special attention was paid to potential theory and
the coupling of equations and different conductors. Thereafter, we
have developed a three dimensional model for internal potential jumps in
electrochemical cells, and implemented it into the finite volume method. 
An internal boundary condition, included into the Laplace
and gradient operator allows for arbitrary floating potential jumps at
electrode-electrolyte interfaces. As it can compute the full
cell potential in 3D at one single mesh, the model is highly efficient,
robust and universal. It is fully parallelized and can work in galvano- and potentiostatic mode. After validating the model
by a simple analytical formula we have illustrated its capabilities by
simulating a Li$||$Bi liquid metal battery and comparing with
experimental data. We have correctly predicted the discharge cycle of
the cell, and illustrated the current and potential distribution as
well as the lithium concentration in bismuth.

Our discretization of the Laplace and gradient terms is quite
universal: only a face field describing the jump needs to be
provided. Our model can therefore be employed for modeling potential
distributions in arbitrary electrochemical cells. Further, it has not escaped
our notice that jumps can appear in electromagnetic fields,
in temperature and potential fields due to contact resistance
\cite{Wang2007} or in concentrations fields at interfaces
\cite{Kaoui2018}, as well. For all those applications, our model might
potentially be useful.

In the future we plan to employ the developed model for studying the
influence of mass transport on the cell potential of liquid metal
batteries \cite{Kelley2018}. Specifically, we would like to extend the
model by convection \cite{Ashour2017a,Weber2018}, heat transfer
\cite{Shen2015,Personnettaz2018a} and electrode kinetics.

\section*{Acknowledgment}
This work was supported by the Deutsche Forschungsgemeinschaft (DFG,
German  Research  Foundation)  by  award  number  338560565  as  well
as a postdoc fellowship of the German Academic Exchange Service (DAAD).
The  computations were performed on a PC-Cluster at the Center for
Information Services and High Performance Computing (ZIH) at TU Dresden
and on the cluster “Hemera” at Helmholtz-Zentrum Dresden -- Rossendorf.
Fruitful discussions with S. Beale, V. Galindo, G. Mutschke and
U. Reimer are gratefully acknowledged. Kashif Mushtaq is grateful to 
the Portuguese Foundation for Science and Technology (FCT) for his 
PhD grant (reference: PD/BD/128041/2016) and support from the MIT 
Portugal Program. 

\bibliography{literature}

\end{document}